\documentclass[runningheads]{llncs}

\usepackage[T1]{fontenc}

\usepackage{graphicx}
\usepackage{array}
\newcolumntype{C}[1]{>{\centering \arraybackslash}m{#1}}

\usepackage{xcolor}
\usepackage{amsmath,amssymb,bbm}
\usepackage{booktabs}
\usepackage{comment}
\usepackage{caption}
\usepackage{subcaption}
\usepackage{svg}
\usepackage{cite}
\newcommand{\TSH}{\rm \left[TSH\right]}
\newcommand{\FT}[1]{\rm \left[FT{#1}\right]}

\begin{document}

\title{Data-based Model Identification of the Hypothalamus-Pituitary-Thyroid Complex}

\titlerunning{Data-based Model Identification of the HPT Complex}

\author{Clara Horvath \and
Andreas Körner \orcidID{0000-0001-7116-1707} \and
Corinna Modiz}

\authorrunning{C. Horvath et al.}

\institute{Institute of Analysis and Scientific Computing, TU Wien,\\
Wiedner Hauptstraße 8-10, 1040 Vienna, Austria \\
\email{corinna.modiz@tuwien.ac.at}}

\maketitle             

\begin{abstract}
The thyroid gland, in conjunction with the pituitary and the hypothalamus, forms a regulated system due to their mutual influence through released hormones.
The equilibrium point of this system, commonly referred to as the “set point”, is individually determined. This means that determining the correct amount of medication to be administered to patients with hypothyroidism requires several treatment appointments creating an extended treatment process.
Because the dynamics of the system have not yet been fully explored, mathematical models are needed to simulate the mutual influence of the respective hormones as well as their course over time. These models enable a deeper understanding of the functionality in the context of data measurements. 
Therefore, two time-dependent mathematical models are presented to replicate this overall influence of disparate systems. Both are based on a system of two differential equations modelling the interacting hormones. 
The parameters of the two models are identified according to different calibration approaches by means of patient data collected in a retrospective study in collaboration with the Medical University of Vienna. 
The hormonal course in the time domain as well as equilibrium curves including the set-point are then simulated and analyzed with respect to the normalized mean squared error. These calibrated systems allow a more profound insight into the functionality of the HPT complex.

\end{abstract}

\section{Introduction}

A prevalent and progressive problem in the general population are thyroid disorders. According to~\cite{Andersen}, the degree of endocrine disorders affecting the thyroid function are between 5\% and 10\% in the general population, therefore research to deepen the understanding of the thyroid gland is becoming increasingly important.

The thyroid gland is an endocrine gland located above the collar bones and below the larynx influencing major physiological functionalities such as metabolism, growth and the cardiovascular system~\cite{Thyroid}.
The functional units of the thyroid gland consists of thyroid follicles responsible for the secretion of the thyroid hormones triiodothyronine (T3), thyroxine (T4), calcitonin and their subsequent release into the bloodstream.
The anterior pituitary regulates the secretion of the first two thyroid hormones through the Thyroid-Stimulating Hormone (TSH). The hypothalamus, in turn, regulates the release of TSH by producing the Thyrotropin-Releasing Hormone (TRH). The Hypothalamus-Pituitary-Thyroid (HPT) complex represents a negative-feedback loop since the production of TSH is suppressed if the concentration of thyroid hormones is high~\cite{Pandiyan}.

Since the mutual influence of the HPT complex is not yet understood to its full extent, simulation of the course of the included hormones based mathematical models is an emerging research field, see~\cite{T1, T2, T3}.
In these models, the hormone TRH is often neglected since accurate data collection is difficult in clinical trials. Thus, hypothalamus and pituitary are combined to form the so-called HP-complex.
Mathematical models often focus on the timeless dynamics of the HPT-complex and, in particular, consider only the mutual influence. 
However, particularly in the context of thyroid disease, time-dependent models are needed to simulate the progression of TSH and FT4 to adjust the corresponding treatment over time. 
Therefore, two time-dependent mathematical models based on a system of differential equations presented in~\cite{Pandiyan, GoedeV1} are analyzed in the context of patient data collected in a retrospective study at Medical University of Vienna. 
The aim of this work is to survey how accurate existing models describe time series of patient measurements and comparing the corresponding results. To achieve this goal, the model parameters are calibrated. 
Additionally, in~\cite{GoedeTheory}, the so-called homeostatic set-point theory is established which introduces a patient-specific pair of TSH and FT4 values in their normal range. 
Thus, the set-point corresponds to the optimal euthyroid state of an individual meaning a healthy condition with respect to the thyroid. 
As the set-point represents the actual target values for both TSH and FT4 of a patient, gathering more information about the dynamics of the HPT-complex in the time domain and the corresponding equilibrium state could lead to better physiological insight and more advanced treatment methods. 

\section{Modelling of the HPT System}

One possible way to obtain a mathematical model describing the interaction and the corresponding control mechanism between the HP complex and the thyroid is to follow the Michaelis-Menten kinetics, known from system biology. The Michaelis-Menten Equation (MME) forms a commonly used model framework for enzymatic reactions, presented in \cite{MME}. The starting point is a system of Ordinary Differential Equations (ODE) representing the dynamics of the enzymatic reaction. A steady-state assumption for the enzyme–substrate complex leads to a simplification of the ODE system and results in the Michaelis-Menten equation of the form 
\begin{align}\label{equ:mmk}
    \frac{\rm d [P]}{{\rm d }t} = \frac{v_\text{max}[S]}{K_M+[S]}.
\end{align}
\noindent
It denotes the product $[P]$ of the enzyme reaction and the substrate $[S]$, the maximum reaction rate $v_\text{max}$ and the so-called Michaelis-Menten constant $K_M$, which describes the reaction rate of an enzymatic reaction. This basic approach, which recalls transfer functions in behavioural modelling, provides the framework for generating dynamic models in systems biology and is a template for the ODE systems used in this paper.

There are several other models describing the HPT system, which are described in~\cite{T1, T2, Liu, Abdu}. This work focuses on two mathematical models using differential equations to describe the steady states of TSH and FT4, i.e., the state in which both hormones stabilize and no longer change their behaviour. 
Both models refer to the hormonal concentration, indicated with $\FT4$ and $\TSH$.
The two models are referred to as model A and model B and will be presented in the following sections.

\subsection{Fundamentals of Model A}

The first model presented in~\cite{GoedeV1} describes the rate of change of the concentrations of $\FT4$ and $\TSH$ in the human blood stream through the two differential equations 
\begin{align}
    \frac{\rm d \TSH}{{\rm d }t} &= k_1 - \frac{k_1 \FT4}{k_a + \FT4} - k_2\TSH, \label{eqn:pandi_tsh}\\
    \frac{\rm d \FT4}{{\rm d}t} &= \frac{k_3 \TSH}{k_d + \TSH} - k_4\FT4, \label{eqn:pandi_ft4}\
\end{align}
\noindent
with initial conditions $\TSH (t_0) = \TSH_0$ and $\FT4(t_0) = \FT4_0$ and the parameters $k_1, k_2, k_3, k_4, k_a, k_d\in \mathbbm{R^+}$. These parameters will be identified on the basis of patient data in later sections. 
The right side of both equations \eqref{eqn:pandi_tsh} and \eqref{eqn:pandi_ft4} represents the secretion rate minus the excretion rate of the respective hormone and the involved fractional term recall the MM kinetics of equation (\ref{equ:mmk}) presented in the previous section. 
The given units for $\TSH$, $\FT4$ and time $t$ are mU/L, pg/mL and days, respectively.

Model A is derived from a system of four differential equations presented in~\cite{Pandiyan}. In addition to the concentration of TSH and FT4, labelled below as $x$ and $y$ respectively, the original model includes equations considering the functional size $z$ of the thyroid gland and anti-thyroid peroxidase antibodies concentration $w$ according to 
\begin{align}
    \frac{{\rm d} x }{{\rm d }t} &= k_1 - \frac{k_1 y}{k_a + y} - k_2x  \label{eqn:old_pandi_tsh},\\
    \frac{{\rm d} y}{{\rm d}t} &= \frac{(k_3 z) x}{k_d + x} - k_4 y \label{eqn:old_pandi_ft4},\\
    \frac{{\rm d} z}{{\rm d} t} &= k_5\left ( \frac{x}{z} - N \right )- k_6 z w, \label{eqn:old_pandi_func} \\
    \frac{{\rm d} w}{{\rm d} t} &= k_7 z w - k_8 w.\label{eqn:old_pandi_AB}
\end{align}
\noindent
The model is represented through the system of differential equations with 11 parameters $k_1$, $k_2$, $k_3$, $k_4$, $k_5$, $k_6$, $k_7$, $k_8$, $k_a$, $k_d$, $N$.
The differential equation \eqref{eqn:pandi_tsh} describing TSH is directly adopted from equation \eqref{eqn:old_pandi_tsh}. Similar, the differential equation for FT4 is mostly identical, however the first term describing the discretion rate of FT4 is slightly modified as the numerator does seemingly not include the the functional size $z$ of the thyroid gland.

In accordance to the normal range of $z$ and the original value of $k_3$ in \eqref{eqn:old_pandi_ft4}, the value of $k_3$ in \eqref{eqn:pandi_ft4} adjusts and incorporates these two aspects in one variable. 
The functional size $z$ is described in the original model though a differential equation, however in model A incorporated in the parameter $k_3$ in \eqref{eqn:pandi_ft4}. Since \eqref{eqn:old_pandi_func} and \eqref{eqn:old_pandi_AB} no longer influence the differential equations for TSH and FT4, consequently they are not longer included. 

\subsection{Fundamentals of Model B}

The second model describing the mutual influence and resulting dynamics of the HPT-complex over time is presented in~\cite{GoedeV1} and further analyzed in~\cite{GoedeTheory, Goede2, Goede3}. 
The system of differential equations including four parameters ${k_1, k_2, k_3, k_4}$ represents the behavior of $\TSH$ over time in negative exponential dependence on $\FT4$ and vice versa, according to

\begin{equation}\label{eqn:geode_ode}
    \begin{split}
    	  \frac{{\rm d}\TSH}{{\rm d}t} &=  \frac{k_1}{\exp(k_2 \FT4)}-\TSH, \\ 
        \frac{{\rm d}\FT4}{{\rm d}t} &=  k_3 - \frac{k_3}{\exp(k_4 \TSH)}-\FT4.
    \end{split} 
\end{equation}

\noindent
This system holds the equations describing its equilibrium state, which means $\tfrac{{\rm d}X}{{\rm d}t} = 0$, with $X\in{\TSH,\FT4}$, leading to

\begin{align} 
	  \TSH &=  \frac{k_1}{\exp(k_2 \FT4)}, \label{eqn:equilibrium_tsh}\\
    \FT4 &=  k_3 - \frac{k_3}{\exp(k_4 \TSH)}. \label{eqn:equilibrium_ft4}
\end{align}

\begin{figure}
\centering
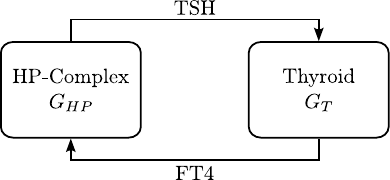
\caption{HPT complex modeled as closed-loop system.} \label{fig:compartment_model}
\end{figure}

\noindent
More specifically, these equations describe the response behavior of the HP-complex and the thyroid, respectively, and can be represented by the corresponding compartments in a control loop, see Fig. \ref{fig:compartment_model}. 

\begin{figure}
\centering
\includegraphics[width=0.7\textwidth]{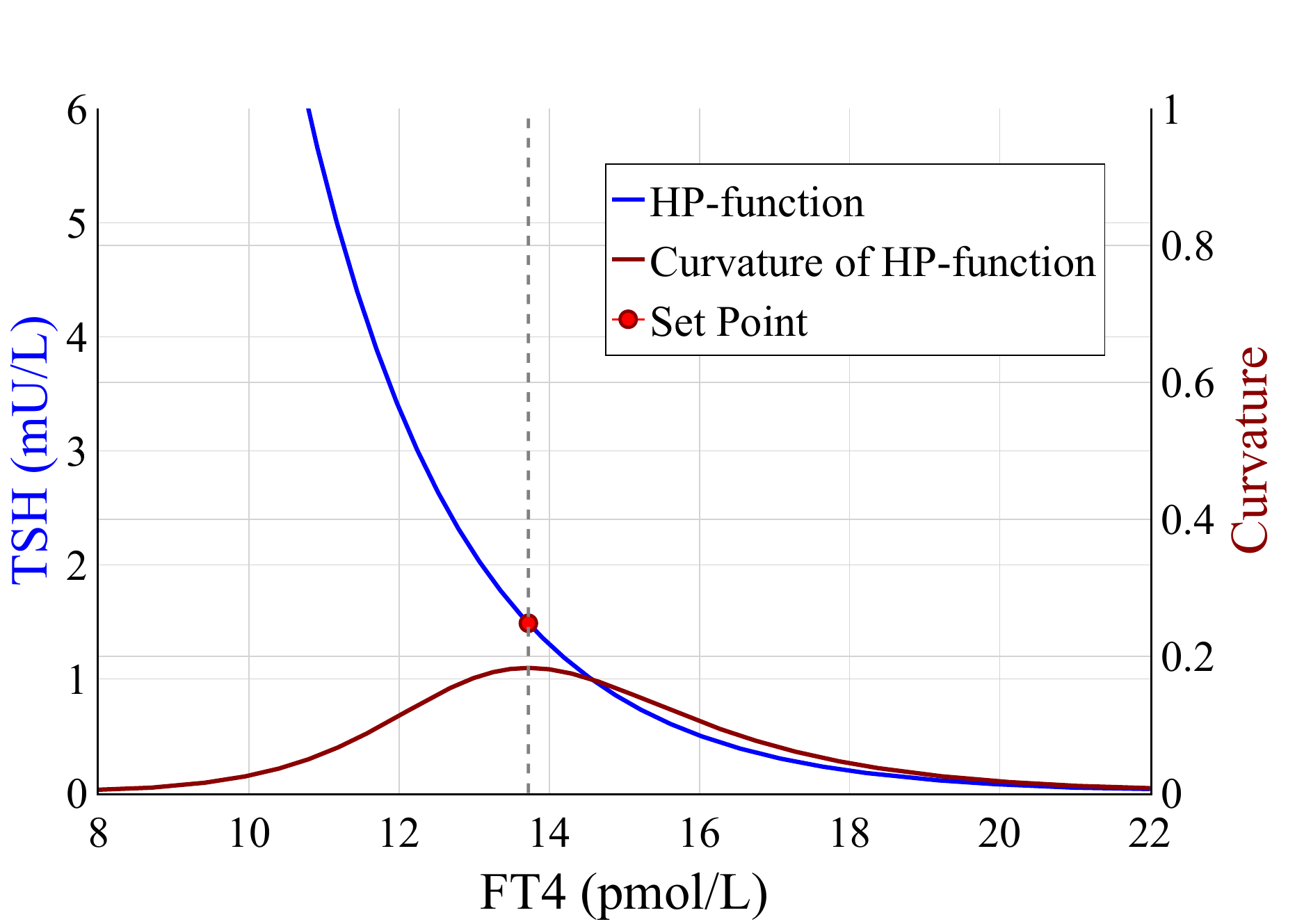}
\caption{HP-function with its set-point defined by the corresponding curvature function.} \label{fig:curvature_curve}
\end{figure}

\newpage
\noindent
Following the theoretical framework in~\cite{GoedeTheory}, the equilibrium equation \eqref{eqn:equilibrium_tsh} of the HP-complex, the so-called HP-function, can be used to compute the set-point.
The set-point $( \TSH_{sp}, \FT4_{sp} )$ represents the specific values of $\TSH$ and $\FT4$ corresponding to a euthyroid state and is patient-specific. It can be expressed explicitly by calculating the point of maximum curvature of the HP-function~\cite{Goede2}, which is analytically proven in~\cite{GoedeTheory} and can be observed in Fig. \ref{fig:curvature_curve}.

\noindent
If this theory is further pursued, an explicit representation for the model parameters $k_3, k_4$ in dependence on the set-point values of $\TSH$ and $\FT4$ can be derived.
The approach is based on the calculation of the optimal closed-loop gain factor $G$  depending on the individual gain factors of the compartment model shown in Fig. \ref{fig:compartment_model}. 

\newpage
\noindent
The optimum of $G$ is found using 

\begin{align} 
        \frac{{\rm d}G}{{\rm d}\TSH} = 0,
\end{align}
with
\begin{align}
	  G =  | G_{HP} \cdot G_{T} |, \quad G_{HP} &= \frac{{\rm d}\TSH}{{\rm d}\FT4}, \quad G_{T} = \frac{{\rm d}\FT4}{{\rm d}\TSH},
\end{align}

\noindent
and the equilibrium equations \eqref{eqn:equilibrium_tsh} and \eqref{eqn:equilibrium_ft4} for $\TSH$ and $\FT4$, respectively. 
As presented in~\cite{GoedeTheory} the constants $k_3$ and $k_4$ can be specified by

\begin{align} 
	  k_3 = \frac{\FT4_{sp}}{\left( 1 - {\rm e}\right)^{-1} } \textrm{\quad and \quad} k_4 = \frac{1}{\TSH_{sp}}. \label{eqn:parameter}
\end{align}

\noindent
Therefore, the parametrization of the differential equation system \eqref{eqn:geode_ode}, can be reduced from four to two parameters $k_1, k_2$.
The approach based on the primarily determination of the set-point may yield different parameters than identifying them using the ODE solution in the time domain.

\section{Methods}

In this section, the data and methods to calibrate the models and simulating the time behaviour of $\TSH$ and $\FT4$ are explained in detail. The general framework for the calibration of the differential equation system is equivalent for both models A and B, but differs in the number of parameters to be determined and in whether the approach takes the time-constrained or the timeless model as the basis for parameterization.
Additionally, the data used for parameterization of the differential equations, is presented shortly.

Both previously presented models are calibrated using a differential evolution algorithm, which is a population-based stochastic algorithm for solving global numerical optimization problems, see~\cite{DiffEvolution}. 
The control variable that specifies the population size was chosen as $10$ to obtain a population consisting of $10$ times the number of dimensions, see~\cite{DiffEvolutionParam}.
To solve the differential equation system, the default explicit Runge-Kutta 45 method is applied. 
The data points of $\TSH$ and $\FT4$ of the respective patients at the first measurement date are chosen as initial condition and the time span is adjusted to represent the entire duration for which measurements of the selected individual are available.
Additionally, the Normalized Mean Squared Error (NMSE) for each curve representing one hormone characteristics over time is calculated 

\begin{align}
\text{error} &=  \text{NMSE}_{\TSH} + \text{NMSE}_{\FT4}, \\
    \text{NMSE}_{X} &= \displaystyle \sum_{i = 1}^N \frac{\left (X(t_i) - \widetilde{X}(t_i)\right )^2 }{N(X_{\rm max} - X_{\rm min})}. 
\end{align}
The variable $\widetilde{X}({t_i})$ represents the simulated hormone concentration using described modules and $X(t_i)$ the corresponding measurement point at time ${t_i}$ for TSH and FT4, respectively. $X_{\rm max}$ is the maximum and $X_{\rm min}$ the minimum of the measured data for each hormone and individual patient. 
It is important to normalize the respective error in order to take into account the different ranges of hormone values of $\TSH$ and $\FT4$. 

\subsection{Data from clinical Studies} \label{sec:data}

The data used for the calibration and analysis of both mathematical models was collected in a retrospective study at Vienna General Hospital in cooperation with the Medical University of Vienna. 
The following calibrations presented in the result section use data from a total of 10 patients, each including three to eight measurements over time.
All of the selected individuals are suffering from a type of hypothyroidism (e.g. autoimmunthyreoditis), including patients receiving a certain dosage of artificially produced T4 hormone starting from a specific time during the measurement. 
The clinical study was originally planned to collect data with respect to thyroid diseases during pregnancy. Therefore, the data includes pregnant patients and measurements on the amount of Human Chorionic Gonadotropin (hCG), see~\cite{HCG}.
One patient is selected for an exemplary presentation of the solutions of the respective models. 
Patient J is pregnant and suffers from hypothyroidism, but is adequately treated with medication throughout her available measurement points.

\subsection{Model-specific Calibration Approaches}

Model A contains six parameters, however not the entire set has been taken into consideration for calibration. As mentioned in~\cite{Pandiyan}, the parameters $k_1$, $k_2$ and $k_4$ are derived from various literature sources and describe physiological constants, and thus are not optimized during calibration. In accordance with~\cite{Pandiyan}, the remaining 3 parameters, namely $k_3$, $k_a$ and $k_d$ are fitted to $\TSH$ and $\FT4$ measurements as these emerge from simulations.

To simulate the course of both hormones over time using model B presented in \eqref{eqn:geode_ode}, two approaches are pursued.
The first approach $\rm B^{sp}$ focuses on the set point derived from the equilibrium equation for $\TSH$ \eqref{eqn:equilibrium_tsh} to determine the parameters $k_1$ and $k_2$. 
According to the previously explained framework, the set-point values of both hormones correspond to the maximum curvature of the calibrated HP-function. 
The remaining parameters $k_3$ and $k_4$ can then be calculated due to the explicit dependency on the set-point given in equation \eqref{eqn:parameter}. 
To describe the timeless course of $\TSH$ as well as $\FT4$ and their mutual influence, the equilibrium equation of $\FT4$ \eqref{eqn:equilibrium_ft4} is inverted resulting in

\begin{align} \label{eq:inverse_ft4}
	  \TSH &=  -\frac{1}{k_4} \log \left(\frac{k_3-\FT4}{k_3} \right).
\end{align}

\noindent
The system of differential equations is then analyzed using the already calibrated, or calculated, parameters in order to simulate the time-constrained course of both hormones depending on patient-data and to determine the respective NMSE. The second approach $\rm B^{ode}$ directly calibrates the system of differential equations \eqref{eqn:geode_ode} in order to fit the two curves to the measurement data in the time domain. 
Subsequently, the error is directly calculated using the NMSE. 

\section{Results}
This section focuses on establishing a relation between the selected methods and the measurement data.
In this context, the determined parameters, the corresponding NMSE and, if applicable, the individual set-point values are presented for each patient. In addition, the resulting hormonal course is shown in the corresponding plots for an exemplary patient described in Section \ref{sec:data}. Although this patient is selected, the hormonal course as a result of both models is similar for all studied patients.

\subsection{Findings in the Course of calibrating Model A}
By simulating model A, the behaviour of the hormone concentration of TSH and FT4 over time are described. 
Depending on the values used in the differential equations \eqref{eqn:pandi_tsh} and \eqref{eqn:pandi_ft4} the corresponding solution and steady state differ.
The solution settles into a stable state after a few days, therefore patient data with well adjusted medication intake are used to calibrate model A, further elaborated in the discussion section. The biologically significant parameters are fixed at $k_1 = 5000$ mU/(L day), $k_2 = 16.63$ 1/day, $k_4 = 0.099$ 1/day, see~\cite{Pandiyan}, while the parameters $k_3$, $k_a$, $k_d$ are calibrated and the results for 10 patients are presented in Table \ref{tab:pandi}. 
\begin{table}[htbp]
\caption{Calibrated parameters including NMSE, mean and standard deviation.\quad} \label{tab:pandi}
    \begin{tabular}{|C{1.cm}|C{2.4cm}|C{2.4cm}|C{2.4cm}|C{2.5cm}|}
\hline
 id &  $k_3$ &  $k_a$ &   $k_d$ &  NMSE \\  
\hline
 A & 1.1947 & 0.2198 & 0.0519 &  0.7744 \\
 B & 1.1350 & 0.1033 & 0.0051 &  0.4930 \\
 C & 1.4096 & 0.0655 & 0.0373 &  0.9185 \\
 D & 0.8905 & 0.0689 & 0.0089 &  0.0754 \\
 E & 1.0663 & 0.0603 & 0.0037 &  0.5783 \\
 F & 1.0932 & 0.0421 & 0.0007 &  0.6368 \\
 G & 1.1262 & 0.0438 & 0.0114 &  0.6004 \\
 H & 1.1949 & 0.1001 & 0.0767 &  0.3980 \\
 I & 0.9519 & 0.0944 & 0.0979 &  0.1232 \\
 J & 1.0981 & 0.0566 & 0.0010 &  0.2752 \\
 \hline
\hline
 mean & 1.1160 & 0.0854 & 0.0294 & 0.4873 \\
 std & 0.1414 & 0.0520 & 0.0351 & 0.2722 \\
 \hline
\end{tabular}
\end{table}

\newpage
\noindent
In comparison with the original model including four differential equations \eqref{eqn:old_pandi_tsh}, \eqref{eqn:old_pandi_ft4}, \eqref{eqn:old_pandi_func}, \eqref{eqn:old_pandi_AB}, the parameter $k_3$ changes significantly in the derived model \eqref{eqn:pandi_tsh}, \eqref{eqn:pandi_ft4}. In the original model a possible value for $k_3$ is given as $86$ pg/mL L day. However, the parameter $k_3$ in the derived model discussed in this work also include the parameter of the functional size $z$, which has a normal range of $0.005 - 0.125$ L, therefore the resulting values of $k_3$ is significantly smaller. Nevertheless, the result for $k_3$ is reasonable, since the functional size $z$ is considered.
\begin{figure}[h!]
     \centering
         \centering
         \includegraphics[width=0.8\textwidth]{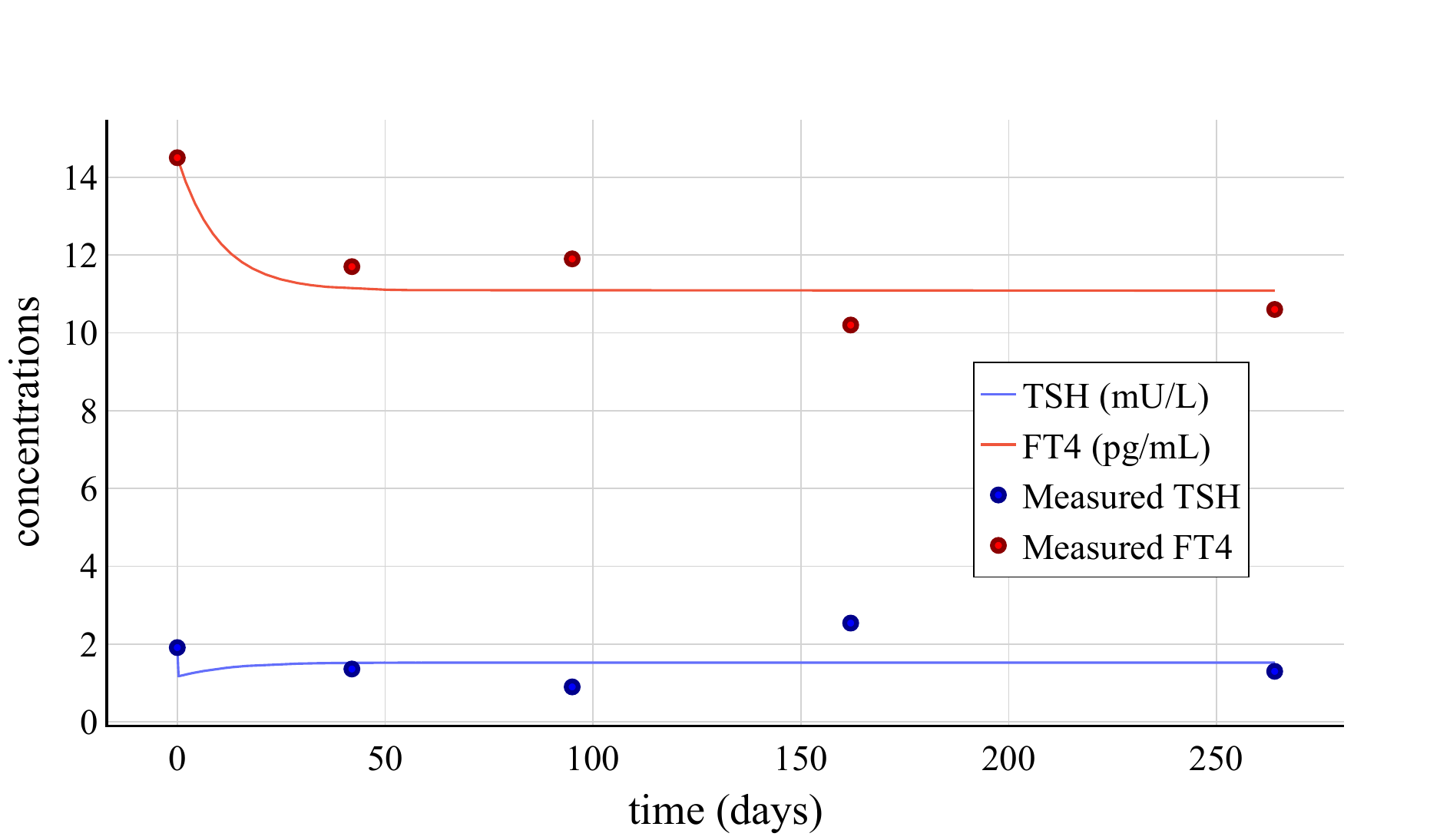}
         \caption{Calibrated solutions of the equations \eqref{eqn:pandi_tsh} and \eqref{eqn:pandi_ft4} for patient J with resulting values $k_3 = 1.0981$, $k_a = 0.0566$, $k_d = 0.001$}
         \label{fig:dgl_65_1}
\end{figure}

\noindent
Both curves in Fig. \ref{fig:dgl_65_1} describing $\TSH$ and $\FT4$ represent the course of the data appropriately and catch the decrease of hormone concentration during the first 50 days. After approximately 60 days, the behaviour of the solution does not change.  
For both curves, the last measured data point is in close proximity to the solution of model A, which suggests that the medication has been adjusted well to fit the steady state of the differential equations \eqref{eqn:pandi_tsh} and \eqref{eqn:pandi_ft4}.

\subsection{Findings in the Course of calibrating Model B}

\begin{figure}[h!]
     \centering
         \centering
         \includegraphics[width=0.7\textwidth]{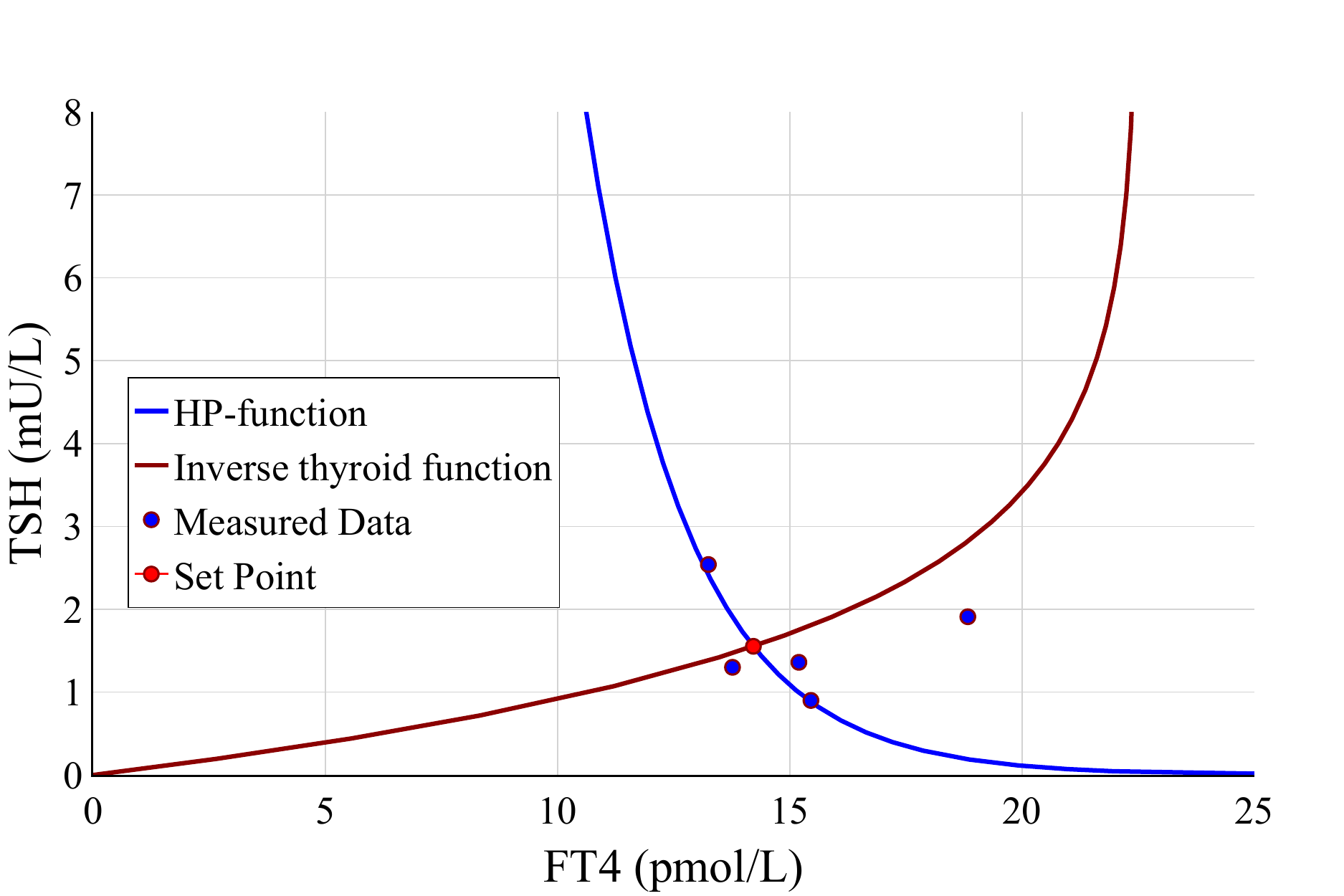}
         \caption{Equilibrium curves, set-point and data of patient J with $\rm FT4_{sp} = 14.22$, $\rm TSH_{sp} = 1.55$, $k_1 = 1002.24$, $k_2 = 0.46$, $k_3 = 22.49$, $k_4 = 0.64$.}
         \label{fig:set_point_65}
\end{figure}

For the approach of simulating the course of the hormones of the HPT complex over time using model B, the parameters of the ODE system \eqref{eqn:geode_ode} are determined in two different ways as discussed in the method section.
First, following approach $\rm B^{sp}$, only the parameters $k_1$ and $k_2$ are determined and the previously explained framework is applied to calculate the set-point as well as the remaining parameters needed among others for the equilibrium equation of the thyroid compartment \eqref{eqn:equilibrium_ft4}.
In order to plot the equilibrium equations \eqref{eqn:equilibrium_tsh}, \eqref{eqn:equilibrium_ft4} along with the corresponding set-point for each data set of a patient, the inverse thyroid function \eqref{eq:inverse_ft4} is selected for the timeless plots. 
In Fig. \ref{fig:set_point_65}, the parameterized HP-function is plotted along with the inverse equilibrium function of the thyroid and the data points of corresponding patient J. 
Both solution graphs intersect at the set-point, which lies within the normal range of FT4 as well as TSH according to~\cite{HormoneRange}.

\begin{figure}[h!]
\centering
\includegraphics[width=0.75\textwidth]{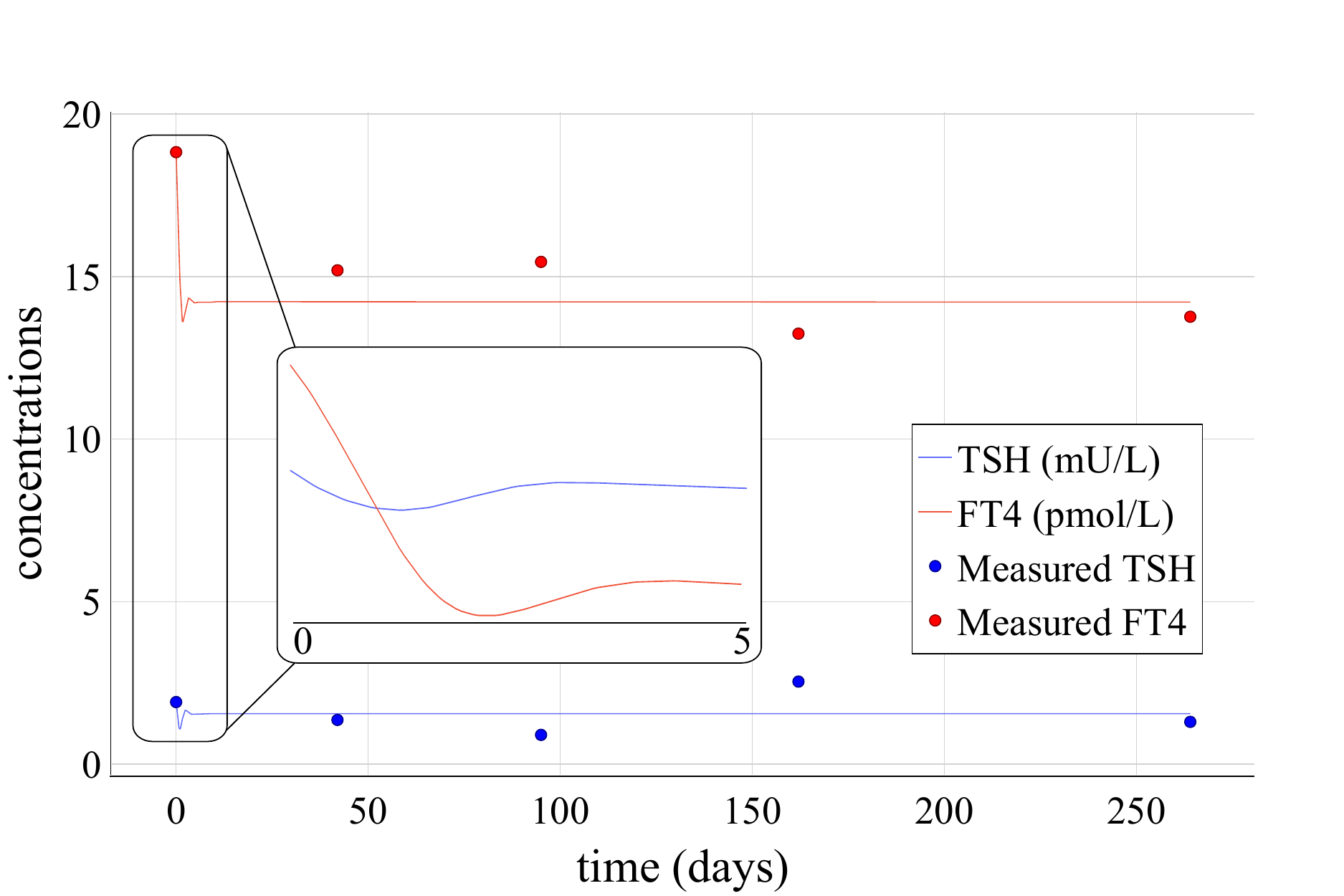}
\caption{Solution curves following approach $\rm B^{sp}$ with $\rm FT4_{sp} = 14.22$, $\rm TSH_{sp} = 1.55$, $k_1 = 1002.24$, $k_2 = 0.46$, $k_3 = 22.49$, $k_4 = 0.64$.}
\label{fig:dgl_65}
\end{figure}

\noindent
Using the calibrated parameters $k_1$, $k_2$, the calculated set-point values and equation \eqref{eqn:parameter}, the course of both hormones over time can be simulated which is presented in Fig. \ref{fig:dgl_65}.

\begin{figure}[h!]
\centering
\includegraphics[width=0.8\textwidth]{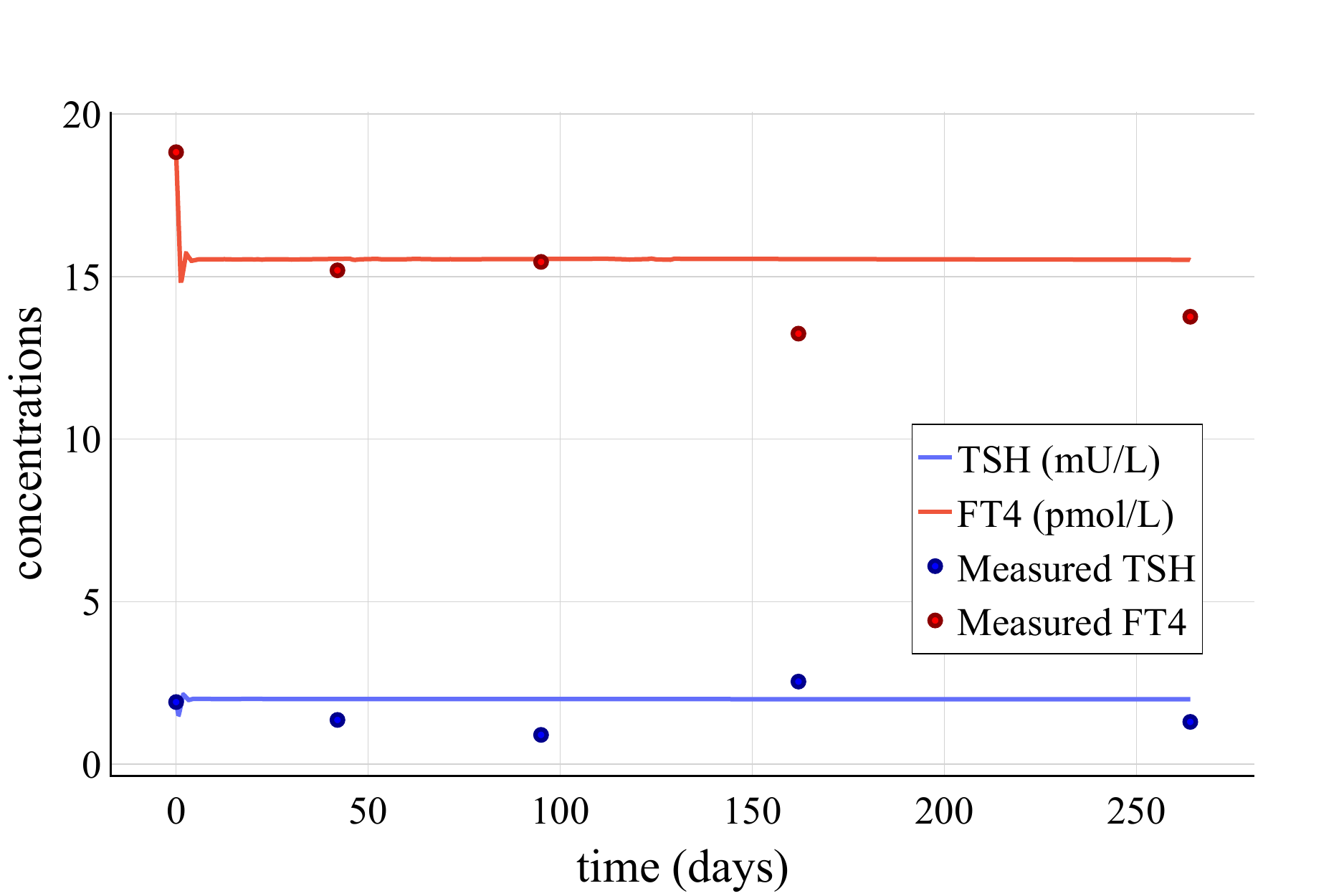}
\caption{Solution curves following approach $\rm B^{ode}$ with $k_1^{\ast} = 1000.40$, $k_2^{\ast} = 0.4$, $k_3^{\ast} =  47.03$, $k_4^{\ast} = 0.2$.}
\label{fig:fit_dgl_65}
\end{figure}

\noindent
Following approach $\rm B^{ode}$, the course of $\TSH$ and $\FT4$ over time resulting from the explicit calibration of both of the solutions of the system \eqref{eqn:geode_ode} to patient-data is shown in Fig. \ref{fig:fit_dgl_65} for the selected patient J.

\noindent
In contrast to approach $\rm B^{sp}$, with approach $\rm B^{ode}$ the parameters $k_1^{\ast}, k_2^{\ast}, k_3^{\ast}, k_4^{\ast}$ are directly calibrated using the solution in the time domain.
The overall results of both approaches are presented in Table \ref{tab:goede}.

\begin{table}[h!]
\caption{Parameters resulting from approach $\rm B^{sp}$ and $\rm B^{ode}$ for 10 patients including set-point, NMSE, mean and standard deviation.}  \label{tab:goede}
\begin{tabular}{|l|l|l|l|l|l|l|l||l|l|l|l|l|l|}
\hline
 id &  $\FT4_{sp}$ &  $\TSH_{sp}$ &   $k_1$ &  $k_2$ &  $k_3$ &  $k_4$ &  NMSE &  $k_1^{\ast}$ &  $k_2^{\ast}$ &  $k_3^{\ast}$ &  $k_4^{\ast}$ &  $\rm NMSE^{\ast}$ \\
\hline
 A &       18.23 &        2.09 & 1002.83 &   0.34 &  28.83 &   0.48 &    5.00 &           1311.66 &              0.35 &             21.94 &              0.23 &               0.85 \\
 B &       16.12 &        1.60 & 1999.68 &   0.44 &  25.50 &   0.63 &    1.77 &           1012.04 &              0.40 &             35.38 &              0.20 &               0.54 \\
 C &       14.59 &        1.60 & 1006.22 &   0.44 &  23.08 &   0.62 &    2.40 &           1390.05 &              0.38 &             77.08 &              0.18 &               1.15 \\
 D &       13.12 &        1.41 & 1000.00 &   0.50 &  20.76 &   0.71 &    0.96 &           1000.27 &              0.40 &             50.59 &              0.11 &               2.05 \\
E &       13.37 &        1.45 & 1001.01 &   0.49 &  21.15 &   0.69 &    0.79 &           1000.13 &              0.40 &             35.88 &              0.24 &               1.23 \\
F &       13.12 &        1.41 & 1000.00 &   0.50 &  20.76 &   0.71 &    1.11 &           1003.70 &              0.40 &             39.03 &              0.29 &               1.39 \\
 G &       13.72 &        1.49 & 1006.84 &   0.47 &  21.71 &   0.67 &    0.92 &           1000.88 &              0.40 &             58.95 &              0.15 &               0.98 \\
 H &       15.74 &        1.75 & 1005.79 &   0.40 &  24.91 &   0.57 &    0.92 &           1106.89 &              0.40 &             52.05 &              0.14 &               0.47 \\
 I &       13.56 &        1.47 & 1000.18 &   0.48 &  21.45 &   0.68 &    3.40 &           1000.17 &              0.40 &             29.69 &              0.19 &               1.57 \\
 J &       14.22 &        1.55 & 1002.24 &   0.46 &  22.49 &   0.64 &    0.31 &           1000.40 &              0.40 &             47.03 &              0.20 &               0.59 \\
\hline
\hline 
mean &  14.58 &  1.58 & 1102.48 &   0.45 &  23.06 &   0.64 &    1.76 &           1082.62 &  0.39 &  44.76 &   0.19 &  1.08 \\
\hline
std &  1.65 &  0.21 & 315.26 &  0.05 &  2.62 &  0.07 &  1.46 &  146.29 & 0.02 &  15.94 &   0.05 &  0.51 \\
\hline
\end{tabular}
\end{table}

\section{Discussion}

The course of the hormonal concentration over time of $\TSH$ and $\FT4$, respectively, is simulated based on two models and, in the case of model B, on two approaches.

Model A represents an abbreviated model of equation \eqref{eqn:old_pandi_tsh}, \eqref{eqn:old_pandi_ft4}, \eqref{eqn:old_pandi_func}, \eqref{eqn:old_pandi_AB}. However, the model shows the evolution of TSH and FT4 concentration and predicts the equilibrium state of these hormones. This information can be used to reflect on whether the resulting equilibrium state is within the normal range of said hormones. 
Therefore, if patients are well-adjusted to their medication, the solution describes measurements, even almost resulting in the last data point, see \ref{fig:dgl_65_1}. With regard to \ref{tab:pandi}, patient A shows interesting properties, as the parameter $k_a$ is significantly larger than any other calibrated value of $k_a$. Furthermore, the NMSE is the second highest. A solution to better approximate the measured data using differential equations, could be the original model as patient A suffers from Hashimoto. Therefore equation \eqref{eqn:old_pandi_AB} could describe the course of anti-thyroid peroxidase antibodies concentration and a overall lower NMSE could be achieved. Further discussion and simulation are needed in order to investigate the effects of reducing the original system of 4 differential equations to model A, which only consists of two equations. Subsequently, a sensitivity analysis is required to better understand the influence of the parameters in model A. Moreover, a significant amount of solution curves of $\FT4$ display a sudden bend after a few days. This phenomenon needs to be further investigated. A possible solution could be using a different numerical solver with adaptive step size. Emerging from this, more information on the dynamic of model A could be derived, since the solution do not reflect the physiological influence between $\TSH$ and $\FT4$.  Concluding, the model A is limited to these data sets exclusively, as the dynamic does not show any oscillating properties. 
Model A is not designed to represent these dynamics, but to represent the development using calibrated parameters. A future goal consists of an improved data approximation which includes the dynamics of said measurements and to propose a corresponding drug administration. 

Model B assumes an exponential mutual dependence of $\TSH$ and $\FT4$ which can be best seen in Fig. \ref{fig:set_point_65}.
This exponential relation differs from the Michaelis-Menten kinetics representing the rational MME, which opens up the potential to investigate this further. 
The HP-function \eqref{eqn:equilibrium_tsh}, calibrated within the framework of approach $\rm B^{sp}$, is derived as equilibrium curve of model B and represents the response behaviour of the HP-complex illustrated in Fig. \ref{fig:compartment_model}. 
As shown in Fig. \ref{fig:set_point_65}, the amount of $\TSH$ released by the pituitary gland is large when the input signal, i.e. the amount of effective $\FT4$, is small and vice versa, which is consistent with the expected physiological behaviour of the HPT complex. 
Also, the response of the thyroid complex, parametrized based on approach $\rm B^{sp}$, is in line with the underlying physiological principal. 
A smaller amount of $\TSH$ as input for the corresponding compartment leads in response to a smaller distribution of $\FT4$ compared to a larger input signal.
The HP function \eqref{eqn:equilibrium_tsh} fits the measurement data of the exemplary patient well and crosses two of the data points almost directly. 
In addition, one point that can be considered an outlier seems to be disregarded for the blue curve, indicating a certain robustness to outliers. One reason for this behaviour could also be the fixed ranges for the calibrated parameters. 
The set point, also included in Fig. \ref{fig:set_point_65}, is determined based on the point of maximum curvature, which is reasonable since it represents the point of highest sensitivity of a function and is therefore consistent with the specificity of that point for each individual patient.
The exemplary depicted set-point is found in a normal range for $\TSH$ and $\FT4$, respectively. 
This indicates that model B, set up to predict the patient-specific set-point, serves its purpose, but it should be noted that the parameters were set to a specific range during calibration, which then in turn influences the course of both curves as well as the set-point values.
In further research, the set-point theory according to~\cite{GoedeTheory} can be compared to the model presented in~\cite{Yang}, that also includes specific values for the hormonal equilibrium described explicitly by a rational function.

The time-dependent solution curves of the approach $\rm B^{sp}$ in Fig. \ref{fig:dgl_65} correspond to the expected physiological behaviour during the first five days, as $\FT4$ decreases with a time delay in response to decreasing $\TSH$. 
Thereafter, an increase in $\FT4$ is observed with a short delay associated with the increase in $\TSH$.
The long-term behaviour shows no dynamics, as the solution curves soon transition to a constant.
One reason for this could be that in~\cite{GoedeV1} the time unit is not explicitly specified, so that the dynamic time span could extend if a different unit is determined. 
Besides, both curves almost meet the last measurement points of patient J. 
This indicates that the equilibrium of system \eqref{eqn:geode_ode} corresponds with her final hormonal values, which are balanced since she is adequately treated with medication according to her description. 
Approach $\rm B^{ode}$ results in solution curves exhibiting the same short and long-term behaviour, shown in Fig. \ref{fig:fit_dgl_65}. 
In contrast, those curves do not meet the last data points as close as approach $\rm B^{sp}$ and the NMSE is almost twice as high, although the curves are directly calibrated to the curve in approach $\rm B^{ode}$.
As listed in Table \ref{tab:goede}, in general, the mean NMSE is smaller for approach $\rm B^{ode}$ compared to $\rm B^{sp}$, which also applies to standard deviation. 
The maximum NMSE for the approach $\rm B^{sp}$ is found for patient A, but she has a fuzzy anamnesis, e.g. she received medication for hyperthyroidism for five appointments before being described medication for hypothyroidism.

Concluding, the dynamic of the solutions of model A and B results in the corresponding steady state of the related differential equations after a relatively short time span, which lies in accordance with the saturation behaviour of MME. Therefore approximating patients' measurements using the presented model is successful for patients with adequately adjusted dosage of medication.
Both models do not reflect any dynamical changes after a certain point in time. However, model B displays the expected physiological behaviour in the time domain as well as in the FT4-TSH plot. 
There are several possible extensions to better reflect the later dynamics of patients measurements, i.e. including the influence of drug administration, which is not included in any of the discussed models.  
Additionally, including the hormone hCG~\cite{HCG} as another input parameter could achieve improved results and broaden the field of application possibilities.

\subsubsection{Acknowledgements} We would like to express our gratitude to our research partner Prof. Dr. Michael Krebs from Medical University Vienna. He provided us with the necessary data of clinical patients. Furthermore, he continuously offered his insight and detailed knowledge of the functionalities of the thyroid gland.


\begin{thebibliography}{10}

\bibitem{Andersen}
Andersen, S., Pedersen, K. M., Bruun, N. H., Laurberg, P.: Narrow Individual Variations in Serum T4 and T3 in Normal Subjects: A Clue to the Understanding of Subclinical Thyroid Disease. The Journal of Clinical Endocrinology and Metabolism \textbf{87}(3), 1088--1072 (2002)

\bibitem{Thyroid}
Yen, P. M.: Physiological and Molecular Basis of Thyroid Hormone Action. Physiological Reviews \textbf{81}(3), 1097--1142 (2001) 

\bibitem{Pandiyan}
Pandiyan, B., Merrill, S. J., Benvenga S.: A patient-specific model of the negative-feedback control of the hypothalamus–pituitary–thyroid (HPT) axis in autoimmune (Hashimoto’s) thyroiditis. Mathematical Medicine and Biology: a journal of the IMA \textbf{31}(3), 226--258 (2013) 

\bibitem{T1}
Berberich, J., Dietrich, J. W., Hoermann, R., Müller, M. A.: Mathematical Modeling of the Pituitary–Thyroid Feedback Loop: Role of a TSH-T3-Shunt and Sensitivity Analysis. Frontiers in Endocrinology \textbf{9}, (2018) 

\bibitem{T2} 
Mukhopadhyay, B., Bhattacharyya, R.: A mathematical model describing the thyroid-pituitary axis with time delays in hormone transportation. Applications of Mathematics \textbf{51}(6), 549--564 (2006) 

\bibitem{T3}
Leow, M. K.: A mathematical model of pituitary–thyroid interaction to provide an insight into the nature of the thyrotropin–thyroid hormone relationship. Journal of Theoretical Biology \textbf{248}(2), 275--287 (2007) 

\bibitem{GoedeV1}
Goede, S. L.: General Review on Mathematical Modeling in the Hypothalamus Pituitary Thyroid System. In: Research Square - Preprint Version 1 (2010). \doi{10.21203/rs.3.rs-1659086/v1}

\bibitem{GoedeTheory}
Leow, M. K., Goede, S. L.: The homeostatic set point of the hypothalamus-pituitary-thyroid axis--maximum curvature theory for personalized euthyroid targets. Theoretical biology \& medical modelling \textbf{11}(1), 35 (2014) 

\bibitem{MME}
Carson, E., Cobelli, C.: Modelling Methodology for Physiology and Medicine. 2nd edn. Elsevier (2014) 


\bibitem{Liu}
Liu, Y., Liu, B., Xie, J., Liu, Y.X.: A new mathematical model of hypothalamo-pituitary-thyroid axis. Mathematical and Computer Modelling \textbf{19}(9), 81--90 (1994)

\bibitem{Abdu}
Abduvaliev, A., Saydalieva, M., Hidirova, M., Gildieva, M.: Mathematical Modeling of the Thyroid Regulatory Mechanisms. American Journal of Medical Sciences and Medicine \textbf{3}(3), 28--32 (2015)

\bibitem{Goede2}
Goede, S. L., Leow, M. K., Smit, J. W. A., Dietrich J. W.: A novel minimal mathematical model of the hypothalamus–pituitary–thyroid axis validated for individualized clinical applications. Mathematical Biosciences \textbf{249}, 1--7 (2014)

\bibitem{Goede3}
Goede, S. L., Leow, M. K.: General Error Analysis in the Relationship between Free Thyroxine and Thyrotropin and Its Clinical Relevance. Computational and Mathematical Methods in Medicine \textbf{2013} 831275 (2014)

\bibitem{DiffEvolution}
Storn, R., Price, K.:  Differential Evolution - A Simple and Efficient Heuristic for Global Optimization over Continuous Spaces. Journal of Global Optimization \textbf{11}(4), 341--359 (1997) 

\bibitem{DiffEvolutionParam}
Centeno-Telleria, M., Zulueta, E., Fernandez-Gamiz, U., Teso-Fz-Betoño, D., Teso-Fz-Betoño, A.:  Differential Evolution Optimal Parameters Tuning with Artificial Neural Network. Mathematics \textbf{9}(4), 424 (2021) 

\bibitem{HormoneRange}
Barth, J., Luvai, A., Jassam, N., Mbagaya, W., Kilpatrick, E., Narayanan, D., Spoors, S.:  Comparison of method related reference intervals for thyroid hormones: studies from a prospective reference population and a literature review. Annals of clinical biochemistry \textbf{55}(1), 107--122 (2017) 

\bibitem{HCG}
Hershman, J.M.: Physiological and pathological aspects of the effect of human chorionic gonadotropin on the thyroid. Best Practice \& Research Clinical Endocrinology \& Metabolism \textbf{18}(2), 249--265 (2004) 


\bibitem{Yang}
Yang, B., Tang, X., Haller, M. J., Schatz, D. A., Rong, L.: A unified mathematical model of thyroid hormone regulation and implication for personalized treatment of thyroid disorders. Journal of Theoretical Biology \textbf{528}, 110853 (2021). 


\end{thebibliography}
\end{document}